\documentclass[aps,twocolumn,showpacs]{revtex4}

\newcommand{\be}{\begin{equation}}
\newcommand{\ee}{  \end{equation}}
\newcommand{\ba}{\begin{eqnarray}}
\newcommand{\ea}{  \end{eqnarray}}
\newcommand{\ve}{\varepsilon}

\begin{document}

\title{Universal Quantum Graphs}

\author{Z. Pluha\v r$^a$ and H. A. Weidenm{\"u}ller$^b$}
\email{Hans.Weidenmueller@mpi-hd.mpg.de}
\affiliation{$^a$Faculty of Mathematics and Physics, Charles University, 180 00 Praha 8, Czech Republic \\ $^b$Max-Planck-Institut f{\"u}r Kernphysik, 69029 Heidelberg, Germany}

\begin{abstract}
For time-reversal invariant graphs we prove the
Bohigas-Giannoni-Schmit conjecture in its most general form: For
graphs that are mixing in the classical limit, all spectral
correlation functions coincide with those of the Gaussian Orthogonal
Ensemble of random matrices. For open graphs, we derive the analogous
identities for all $S$-matrix correlation functions.
\end{abstract}

\maketitle

{\it Introduction}. The distribution of eigenvalues of a classically
chaotic Hamiltonian is one of the central themes of quantum chaos. In
1984, Bohigas, Giannoni, and Schmit (BGS)~\cite{Boh84} formulated the
celebrated ``BGS conjecture'' (see also Refs.~\cite{McD79, Cas80,
  Ber81}): {\it The spectral fluctuation properties of a Hamiltonian
  quantum system that is classically chaotic (mixing) coincide with
  those of the random-matrix ensemble in the same symmetry class.}
Here ``spectral fluctuation properties'' refers to the totality of
spectral fluctuation measures. The symmetry class (orthogonal,
unitary, or symplectic) is determined~\cite{Dys62} by the properties
of the system under time reversal and under rotation.

In addition to substantial numerical evidence~\cite{Haa10}, the BGS
conjecture has received analytical support along two lines. (i) With
the help of the semiclassical approximation and periodic-orbit theory,
the level-level correlator (``two-point function'') for chaotic
systems was shown to coincide with that of random-matrix
theory~\cite{Mue04, Mue05, Heu07, Heu09}. (ii) The two-point function
for quantum graphs~\cite{Kot99} was shown~\cite{Gnu04, Gnu05} to obey
the BGS conjecture (even though graphs are not strictly Hamiltonian
systems). That result was extended to the $S$-matrix correlation
function for open graphs~\cite{Plu13a, Plu13b}.

In this Letter we prove the BGS conjecture for time-reversal invariant
graphs in its most general form. Generalizing the approach of
Refs.~\cite{Gnu04, Gnu05, Plu13a, Plu13b} we show that for graphs with
incommensurate bond lengths that are mixing in the classical limit,
all spectral correlation functions coincide with those of the
Gaussian Orthogonal Ensemble (GOE) of random matrices. For open
graphs, we derive the analogous identities for all $S$-matrix
correlation functions.

{\it Graphs}. We need to define the correlation functions for levels
and for $S$-matrix elements. To make the paper self-contained we first
collect the relevant definitions and properties of graphs.  A closed
graph~\cite{Kot99, Gnu06} is a system of $V$ vertices labelled
$\alpha, \beta, \ldots$ connected by $B$ bonds labelled $(\alpha
\beta), \ldots$ or simply by $b = 1, 2, \ldots, B$. We consider
simple, completely connected graphs (every pair of vertices is
connected by a single bond). Then $B = V (V - 1) / 2$. We eventually
take the limit $B \to \infty$.  The lengths $L_b$ of the bonds are
assumed to be incommensurate (there is no set $\{ i_b \}$ of positive,
negative or zero integers for which $\sum_b i_b L_b$ vanishes). For $B
\to \infty$, the lengths are assumed to remain bounded, $L_{\rm min}
\leq L_b \leq L_{\rm max}$ for all $b$. On each bond $b$ the
Schr{\"o}dinger wave is written as $s_{b 1} \exp \{ i k x_b \} + s_{b
  2} \exp \{ - i k x_b \}$ with the same real wave number $k$ for all
bonds. The variable $x_b$ denotes the distance to one of the two
vertices connected by the bond. The set of coefficients $\{ s_{b 1},
s_{b 2} \}$ is determined by boundary conditions defined on each
vertex $\alpha$ and written as ${\cal O}^{(\alpha)} = \sigma^{(\alpha)}
{\cal I}^{(\alpha)}$. Here ${\cal I}^{(\alpha)}$ (${\cal O}^{(\alpha)}$) is
the vector of incoming (outgoing) wave amplitudes on the bonds
attached to vertex $\alpha$, respectively. The matrices
$\sigma^{(\alpha)}$ have dimension $V - 1$ and are symmetric
(time-reversal invariance) and unitary (flux conservation). Open
graphs are defined by attaching to each of the vertices labeled
$\alpha = 1, 2, \ldots, \Lambda$ an additional bond (a ``channel'')
labeled $\alpha$ that extends to infinity. For these vertices the
boundary conditions ${\cal O}^{(\alpha)} = \Gamma^{(\alpha)} {\cal
  I}^{(\alpha)}$ involve the symmetric and unitary boundary condition
matrices $\Gamma^{(\alpha)}$ of dimension $V$ given by
\be
\Gamma^{(\alpha)} = \left( \matrix{
\rho^{(\alpha)} & \tau^{(\alpha)}_\beta \cr
\tau^{(\alpha)}_\gamma & \sigma^{(\alpha)}_{\gamma \beta} \cr} \right) \ .
\label{1}
\ee
Here $\rho^{(\alpha)}$ is the amplitude for backscattering into
channel $\alpha$, $\tau^{(\alpha)}_\beta$ is the amplitude for
scattering from channel $\alpha$ to vertex $\beta$ or vice versa. The
matrices $\sigma^{(\alpha)}$ in Eq.~(\ref{1}) are subunitary. For $B
\to \infty$, the number $\Lambda$ of channels is held fixed.

To introduce the spectral determinant $\xi(k)$ for closed
graphs~\cite{Kot99, Gnu04, Gnu05, Gnu06} and the scattering matrix
($S$ matrix) $S(k)$ for open graphs~\cite{Kot99, Plu13a, Plu13b}, we
define in both cases the block-diagonal symmetric ``vertex scattering
matrix'' $\Sigma^{(V)}$.  That matrix contains the matrices
$\sigma^{(\alpha)}$, $\alpha = 1, 2, \ldots, V$ in its diagonal
blocks. It has dimension $V (V - 1)$ and is unitary (subunitary) for
closed (open) graphs, respectively. Since $V (V - 1) = 2 B$, all
relevant expressions can most easily be written in matrix form by
doubling the number of bonds. The bonds $(\alpha \beta)$ are arranged
in lexicographical order (so that $\alpha < \beta$). The resulting
sequence is mapped onto the sequence $b = 1, \ldots, B$. These bonds
carry the labels $(b +)$. To every such ``directed bond'' $(\alpha
\beta)$ with $\alpha < \beta$ the bond $(b -)$ is defined by $(\beta
\alpha)$. The number of directed bonds $(b d)$ with $d = \pm$ is $2
B$. In directed bond representation, the vertex scattering matrix is
denoted by $\Sigma^{(B)}$ (``bond scattering matrix''). That matrix is
also symmetric and unitary (subunitary, respectively).

Zeros of the spectral determinant $\xi(k) = \det (1 - \exp \{ i k
{\cal L} \} \sigma^d_1 \Sigma^{(B)})$ define the bound states of a
closed graph while scattering on an open graph is described by the
symmetric unitary scattering matrix $S_{\alpha \beta}(k)$ of dimension
$\Lambda$,
\be
S_{\alpha \beta}(k) = \rho^{(\alpha)} \delta_{\alpha \beta} + \big( {\cal T}
{\cal W}^{- 1} {\cal T}^T \big)_{\alpha \beta} \ .
\label{2}
\ee
Here ${\cal W} = \exp \{ - i k {\cal L} \} \sigma^d_1 - \Sigma^{(B)}$
while ${\cal T}$ is a rectangular matrix of dimension $\Lambda \times
2 B$ containing the amplitudes $\tau^{(\alpha)}_\beta$ in directed-bond
representation as non-zero elements. The symbol $T$ denotes the
transpose. The matrix $\exp \{ i k {\cal L} \}$ with ${\cal L} = \{
\delta_{b b'} \delta_{d d'} L_b \}$ describes propagation on the
directed bonds, with the bond propagator $\exp \{ i k L_b \}$
independent of the direction of the bond. The matrix $\sigma^d_1$ is
the first Pauli spin matrix in directional space multiplied by the
unit matrix in non-directed bond space. That matrix is needed to write
$\xi(k)$ and $S_{\alpha \beta}(k)$ in matrix form.

The probability distributions for levels and $S$-matrix elements are
specified in terms of average values and correlation functions. All
averages (indicated by angular brackets) are taken over the wave
number $k$. The average level density is~\cite{Kot99} $\langle d_{\rm
  R} \rangle = (1 / \pi) \sum_b L_b$, the average $S$ matrix
is~\cite{Kot99} $\langle S_{\alpha \beta} \rangle = \rho^{(\alpha)}
\delta_{\alpha \beta}$. The fluctuating part of the level density
is~\cite{Gnu04, Gnu05} $(1 / (i \pi)) \Im ({\rm d} / ({\rm d} k)) \ln
\xi(k^+)$ where the plus sign indicates an infinitesimal positive
imaginary increment. The fluctuating part of the scattering matrix is
$S^{\rm fl} = {\cal T} {\cal W}^{- 1} {\cal T}^T$. In terms of these
quantities, the $(P, Q)$ correlation functions for levels (closed
graphs) and $S$-matrix elements (open graphs) are
\ba
&& \bigg\langle \prod_{p = 1}^P \frac{\rm d}{{\rm d} k} \ln \xi(k^+ +
\kappa_p) \prod_{q = 1}^Q \frac{\rm d}{{\rm d} k} \ln \xi(k^- -
\tilde{\kappa}_q) \bigg\rangle \ , \nonumber \\
&& \bigg\langle \prod_{p = 1}^P S^{\rm fl}_{\alpha_p \beta_p}(k +
\kappa_p) \prod_{q = 1}^Q S^{{\rm fl} *}_{\gamma_q \delta_q}(k -
\tilde{\kappa}_q) \bigg\rangle \ .
\label{3}
\ea
Here $P$ and $Q$ are positive integers. Interest centers on
fluctutions on the scale of the average level spacing $1 / \langle
d_{\rm R} \rangle$. Therefore, the incremental wave numbers $\kappa_p$
and $\tilde{\kappa}_q$ obey $\kappa_p \langle d_{\rm R} \rangle \ll B$
and $\tilde{\kappa}_q \langle d_{\rm R} \rangle \ll B$.

{\it Classical Limit}. In the classical limit~\cite{Bar01, Pak01} the
time evolution of the probability density $r$, a vector with component
$r_{b d} \geq 0$ on the directed bond $(b d)$, is defined in terms of
the discrete map $r \to {\cal F} r$. Here ${\cal F}$ is the
Perron-Frobenius (PF) operator~\cite{Gnu06}, a non-symmetric matrix
with elements ${\cal F}_{b d, b' d'} = |(\sigma^d_1 \Sigma^{(B)})_{b
  d, b' d'}|^2$. All elements of ${\cal F}$ are positive or
zero. Moreover, $\sum_{b' d'} {\cal F}_{b d, b' d'} = 1 = \sum_{b d}
{\cal F}_{b d, b'd'}$. According to the Perron-Frobenius theorem,
${\cal F}$ possesses a non-degenerate maximal eigenvalue $\lambda_1 =
1$ with associated right (left) eigenvector $u_1 = (1, 1, \ldots,
1)^T$ ($w_1 = (1, 1, \ldots, 1)$) respectively. A closed graph is
mixing~\cite{Gnu06} if $\lambda_1$ is the only eigenvalue of ${\cal
  F}$ on the unit circle in the complex plane, with all other
eigenvalues located within or on the surface of a disc within the unit
circle. For an $m$-fold repeated map $r \to {\cal F}^m r$ we then have
$r \to u_1 ( w_1 | r )$ exponentially for $m \to \infty$. For the
graph to remain mixing in the limit $B \to \infty$ we require that the
minimum distance between the disc of eigenvalues $\lambda_j$ with $j
\geq 2$ and the unit circle remains finite, $|\lambda_j| \leq 1 - a$
with $a > 0$. We postulate that same condition for open graphs. Since
$\Lambda$ is held fixed it seems plausible that for $B \to \infty$,
that condition is met (perhaps with a different value of $a$) for any
open graph the closed counterpart of which is mixing.

{\it Averages. Supersymmetry}. To average over $k$, the content of the
angular brackets in every $(P, Q)$ correlation function is written as
a suitable derivative of a generating function ${\cal G}_G$ (a
superintegral)~\cite{Gnu05, Plu13b}. The average is carried out using
the incommensurability of the bond lengths by replacing~\cite{Gnu04}
the integration over $k$ by an integration over the independent phases
$\phi_b = k L_b$ and using the color-flavor
transformation~\cite{Zir96}. The $P$ ($Q$) factors in
expressions~(\ref{3}) generate the retarded block (the advanced block,
respectively). The result is~\cite{Gnu05, Plu13b} $\langle {\cal G}_G
\rangle$ $= \int {\rm d} (Z, \tilde{Z})$ $\exp \{ - {\cal A} \}$ where
the effective action is
\ba
&& {\cal A}(Z, \tilde{Z}) = - {\rm STL} ( 1 - Z \tilde{Z}) +
\frac{1}{2} {\rm STL} ( 1 - z_+ Z z_- Z^\tau) \nonumber \\
&& + \frac{1}{2} {\rm STL} (1 - z_+ S_+ \tilde{Z}^\tau S_-^\dag z_-
\tilde{Z}) \ .
\label{4}
\ea
Here ${\rm STL}$ stands for the combined operations $({\rm STr}
\ \ln)$ where ${\rm STr}$ denotes the supertrace. Moreover, $S_\pm =
(\sigma^d_1 \Sigma^{(B)} - {\cal J}_\pm)$ while $z_+ = \exp \{ i
\kappa {\cal L} \}$ and $z_- = \exp \{ i \tilde{\kappa} {\cal L} \}$
in obvious notation. Differentiation of ${\cal G}_G$ with respect to
the source terms ${\cal J}_+$ (${\cal J}_-$) in the retarded
(advanced) blocks yields the $(P, Q)$ correlation functions. The
source terms ${\cal J}_\pm$ differ for open and for closed graphs and
are given in Refs.~\cite{Gnu04, Gnu05, Plu13a, Plu13b}. With $s = 1,
2, 3, 4$ the index for the supervariables, the matrices $Z$
($\tilde{Z}$) have elements $Z_{p b d s, q b' d' s'}$ ($\tilde{Z}_{q b
  d s, p b' d' s'}$), dimension $8 B P \times 8 B Q$ ($8 B Q \times 8
B P$, respectively), and are both diagonal ($\propto \delta_{b b'}$)
in bond space. The integration measure is the flat Berezinian. As in
Ref.~\cite{Gnu05} $Z^\tau$ is a transform of $Z$.

{\it Saddle-Point Manifold}. Variation of ${\cal A}(Z, \tilde{Z})$
with respect to $Z$ and $\tilde{Z}$ yields two saddle-point
equations~\cite{Gnu04, Plu13b}. The first one yields $Z^\tau =
\tilde{Z}$. The second is met if (i) $[\sigma^d_1 \Sigma^{(B)} , Z] =
0$ and if (ii) $\sigma^d_1 \Sigma^{(B)} (\sigma^d_1 \Sigma^{(B)})^\dag
= 1$. Condition (i) reduces the matrices $Z$, $\tilde{Z}$ to the
saddle-point solution $Y = \{ \delta_{b b'} \delta_{d d'} Y_{p s, q
  s'} \}$, $\tilde{Y} = \{ \delta_{b b'} \delta_{d d'} \tilde{Y}_{q s,
  p s'} \}$. In saddle-point approximation we have $\langle {\cal G}_G
\rangle_{s p}$ $= \int {\rm d} (Y, \tilde{Y})$ $( \ldots) \exp \{ SB_G
+ CC_G \}$ where the integration measure is the flat Berezinian. The
dots indicate the source terms. The ``symmetry-breaking term'' is
\be
SB_G = i \pi \langle d_{\rm R} \rangle \bigg\{ \sum_p \kappa_p
{\rm STr}_s \bigg( \frac{1}{1 - Y \tilde{Y}} \bigg)_{p p} + \ldots
\bigg\} \ .
\label{5}
\ee
The dots indicate a second term obtained from the first by $p \to q$,
$\kappa_p \to \tilde{\kappa}_q$, $Y \leftrightarrow \tilde{Y}$.
Condition (ii) is violated for open graphs. The unitarity deficit of
the matrices $\sigma^{(\alpha)}$ for $\alpha = 1, \ldots, \Lambda$ and
the ensuing unitarity deficit of the average $S$ matrix are accounted
for by the ``channel-coupling term''
\be
CC_G = - \frac{1}{2} \sum_{\alpha = 1}^\Lambda
      {\rm STr}_{p s} \ln \bigg( 1 + T^{(\alpha)} \frac{Y \tilde{Y}}{1
        - Y \tilde{Y}} \bigg) \ .
\label{6}
\ee
where $T^{(\alpha)} = 1 - |\langle S_{\alpha \alpha} \rangle |^2$.

{\it Massive Modes}. The degrees of freedom in the superintegral for
$\langle {\cal G}_G \rangle$ that do not belong to the saddle-point
manifold are orthogonal to $Y$, $\tilde{Y}$ and are taken into account
in Gaussian approximation~\cite{Gnu05}. We expand the effective
action~(\ref{4}) up to second order in $Z$, $\tilde{Z}$, dropping the
source terms and the incremental wave numbers $\kappa_p$,
$\tilde{\kappa}_q$. We use $Z = \tilde{Z}^\tau$ since fluctuations
away from that condition are strongly suppressed~\cite{Gnu05}. That
yields two terms. One contains $(1 - {\cal F})_{b d, b' d'}$
sandwiched between $Z_{p b d s, q b d s'}$ and $ \tilde{Z}_{q b' d' s,
  p b' d' s'}$. It can be written as
\be
\frac{1}{2} \sum_{p = 1}^P \sum_{q = 1}^Q \sum_{j \geq 2}^{2 B} {\rm STr}_s
\bigg\{ z_{j; p q} ( 1 - \lambda_j ) \tilde{z}_{j; q p} \bigg\} \ .
\label{7}
\ee
The supermatrices $z_{j; p q}$ ($\tilde{z}_{j; q p}$) are obtained by
multiplying $Z_{p b d s; q b d s'}$ ($\tilde{Z}_{q b' d' s; p b' d'
  s'}$) with the left (right) eigenvectors of ${\cal F}$,
respectively, that belong to eigenvalue $\lambda_j$ with $j \geq 2$.
Since $\Re \lambda_j < 1$ for all $j \geq 2$, the expression~(\ref{7})
defines {\it bona fide} Gaussian integrals with masses $m_j = 1 -
\lambda_j$ for $j \geq 2$, both for closed and for open graphs. The
second term is the supertrace of $[1 - (\sigma^d_1 \Sigma^{(B)})_{b d,
    b d} (\sigma^d_1 \Sigma^{(B)})^\dag_{b d', b d'}]$ sandwiched
between $Z_{p b d s, q b d' s'}$ and $\tilde{Z}_{q b d' s, p b d s'}$
and summed over all $b$ and all $d \neq d'$. The fluctuations due to
that term are negligible because the matrices $\sigma^{(\alpha)}$ are
unitary or subunitary so that for $V \gg 1$ all elements of
$\sigma^{(\alpha)}$ are generically small (of order $V^{- 1/2}$). We
focus attention on expression~(\ref{7}). We expand the source terms
and the remaining terms in the effective action~(\ref{4}) in Taylor
series in $Z_{p b d s, q b d s'}$ and $\tilde{Z}_{q b' d' s, p b' d'
  s'}$, dropping all other terms.  Using the right and left
eigenfunctions of ${\cal F}$ we transform $Z_{p b d s, q b d s'} \to
z_{j; p s, q s'}$ ($\tilde{Z}_{q b d s, p b d s'} \to z_{j; q s, p
  s'}$, respectively). We carry out the resulting Gaussian
integrals. For closed graphs, the resulting expressions are bounded
from above by terms of the form
\be
\frac{C}{B} \prod_{l = 1}^{P + Q - 1} \frac{1}{B} \sum_{j_l = 2}^{2 B}
\frac{1}{|m_{j_l}|^{k_l}} \ .
\label{8}
\ee
Here $C$ is some positive constant and $k_l$ are non-negative
integers. For $|m_j| > a$ (all $j$) the term~(\ref{8}) vanishes for $B
\to \infty$. The factors $B^{- 1}$ in expression~(\ref{8}) are due to
the source terms for closed graphs. Detailed analysis shows that
reduction factors equivalent to $B^{- 1}$ arise also for open graphs
because here the source terms are matrices in directed bond space that
have a single nonvanishing element only. Hence, the expressions
analogous to~(\ref{8}) for open graphs also vanish.

We conclude that both for closed and for open graphs, the contribution
of massive modes is negligible for $B \to \infty$. Therefore, all $(P,
Q)$ correlation functions are obtained by differentiating $\langle
{\cal G}_G \rangle_{s p}$ with respect to the source terms.

{\it Random-Matrix Approach}. We turn to the GOE~\cite{Meh04} and
generalize the supersymmetry approach of Refs.~\cite{Efe83, Ver85} to
the general $(P, Q)$ correlation function. The real matrix elements
$H_{\mu \nu}$ of the symmetric $N$-dimensional GOE Hamiltonian $H$ are
Gaussian-distributed random variables with zero mean values and second
moments $\langle H_{\mu \nu} H_{\mu' \nu'} \rangle = (\lambda^2 / N)
(\delta_{\mu \mu'} \delta_{\nu \nu'} + \delta_{\mu \nu'} \delta_{\nu
  \mu'})$. The indices run from $1$ to $N$ while $\lambda = N d / \pi$
where $d$ is the mean level spacing at the center of the GOE
spectrum. The angular brackets denote the ensemble average. With $E$
the energy, the $(P, Q)$ level correlation function for the closed
system is defined as
\be
\bigg\langle \prod_{p = 1}^P {\rm Tr} (E^+ + \ve_p - H)^{- 1}
\prod_{q = 1}^Q {\rm Tr} (E^- - \tilde{\ve}_q - H)^{- 1} \bigg\rangle \ .
\label{9}
\ee
The plus (minus) sign indicates an infinitesimal positive (negative)
imaginary increment. The open system is obtained~\cite{Ver85} by
coupling $\Lambda$ channels $a, b, \ldots$ to the states labeled $\mu$
by real channel-coupling matrix elements $W_{a \mu} = W_{\mu
  a}$. These obey $\sum_\mu W_{a \mu} W_{\mu b} = N v^2_a \delta_{a
  b}$. The scattering matrix is $S_{a b} = \delta_{a b} - 2 \pi i [W
  (E - H + i \pi W^\dag W)^{- 1} W^\dag]_{a b}$.  The $S$-matrix
correlation function is defined in analogy to the second term of
expression~(\ref{3}), with the replacements $S^{\rm fl}_{\alpha_p
  \beta_p}(k + \kappa_p) \to S_{a_p b_p}(E + \ve_p)$, $S^{\rm fl
  *}_{\gamma_q \delta_q}(k - \tilde{\kappa}_q) \to S^*_{c_q d_q}(E -
\tilde{\ve}_q)$. In contrast to expression~(\ref{3}) the correlator
now also contains the average $S$-matrix elements. That must be borne
in mind when we later compare the source terms. The incremental
energies obey $\ve_p, \tilde{\ve}_q \ll d N$.

The contents of the angular brackets in the $(P, Q)$ correlation
functions are written as suitable derivatives~\cite{Ver85a, Ver85}
with respect to source terms ${\cal J}_\pm$ of a generating function
${\cal G}_R$ (a superintegral). The ensemble average is calculated by
straightforward generalization of the steps in Ref.~\cite{Ver85}.  The
ensemble average over $H$ is followed by the Hubbard-Stratonovich
transformation and by the saddle-point approximation. At the center of
the GOE spectrum, the saddle-point manifold is parametrized as
$\sigma_R = - i T^{- 1}_0 L T_0$. In retarded-advanced block notation
$L$ is equal to the third Pauli spin matrix while $T_0$ is given by
\be
T_0 = \left( \matrix{ (1 + t_{1 2} t_{2 1})^{1/2} & i t_{1 2} \cr
         - i t_{2 1} & (1 + t_{2 1} t_{1 2})^{1/2} \cr} \right) \ . 
\label{10}
\ee
The matrix $t_{1 2}$ ($t_{2 1}$) has elements $(t_{1 2})_{p s, q s'}$
($(t_{2 1})_{q s, p s'}$, respectively). The elements of $(t_{1 2},
t_{2 1})$ span the saddle-point manifold for the $(P, Q)$ correlation
function. That gives $\langle {\cal G}_R \rangle_{s p} = \int {\rm d}
\mu(t) ( \ldots ) \exp \{ SB_R + CC_R \}$ where the dots indicate the
source terms. We suppress the definition of the invariant measure
${\rm d} \mu(t)$. In analogy to Eqs.~(\ref{5}, \ref{6}) the
symmetry-breaking term is
\be
SB_R = {i \pi \over d} \bigg\{ \sum_p \ve_p {\rm STr}_s \bigg(
(t_{1 2} t_{2 1})_{p p} \bigg) + \ldots \bigg\}
\label{11}
\ee
where the dots indicate a second term obtained from the first by the
replacements $p \to q$, $\ve_p \to \tilde{\ve}_q$, $(t_{1 2} t_{2
  1})_{p p} \to (t_{2 1} t_{1 2})_{q q}$). The channel-coupling
term (present only for the open system) is
\be
CC_R = - {1 \over 2} \sum_c {\rm STr}_{p s} \ln \bigg(1 + T^{(c)} t_{1 2}
t_{2 1} \bigg) \ .
\label{12}
\ee
The transmission coefficient $T^{(c)}$ in channel $c$ is defined as
$T^{(c)} = 1 - |\langle S_{c c} \rangle|^2$.

The contribution of the massive modes to the $(P, Q)$ correlation
functions for the GOE can be shown to vanish with some inverse power
of $N$ as $N \to \infty$. Therefore, these functions are obtained by
differentiation of $\langle {\cal G}_R \rangle_{s p}$ with respect to
the sources.

{\it Equivalence}. For $B \to \infty$ and $N \to \infty$, massive
modes contribute neither to $\langle {\cal G}_G \rangle$ nor to
$\langle {\cal G}_R \rangle$. The identity of all $(P, Q)$ correlation
functions of both approaches is, therefore, proved by showing that
$\langle {\cal G}_G \rangle_{s p} = \langle {\cal G}_R \rangle_{s p}$. We
equate $\ve_p / d$ with $\kappa_p \langle d_{\rm R} \rangle$,
$\tilde{\ve}_q / d$ with $\tilde{\kappa}_q \langle d_{\rm R} \rangle$,
$T^{(a)}$ with $T^{(\alpha)}$ for both $a$ and $\alpha = 1, \ldots,
\Lambda$. We define
\be
\tau = - i t_{1 2} \frac{1}{\sqrt{1 + t_{2 1} t_{1 2}}} \ , \
\tilde{\tau} = i t_{2 1} \frac{1}{\sqrt{1 + t_{1 2} t_{2 1}}} \ .
\label{13}
\ee
With these substitutions and upon the identification $\tau = Y$,
$\tilde{\tau} = \tilde{Y}$, the terms $SB_R$ and $CC_R$ in
Eqs.~(\ref{11}) and (\ref{12}) become equal to $SB_G$ and $CC_G$ in
Eqs.~(\ref{5}) and (\ref{6}), respectively. For the source terms (not
given here) the identity is easily proved for the closed systems. For
the open systems, the identity is established on the level of the
transmission coefficients as the coupling matrix elements $W_{a \mu}$
of the GOE approach bear no direct analogy to the elements of the
matrix $\Sigma^{(B)}$ for graphs.

With the substitutions~(\ref{13}) the saddle-point manifold $\sigma_R
= - i T^{- 1}_0 L T_0$ takes the form
\be
\sigma_R = - i \left( \matrix{
1 & \tau \cr
\tilde{\tau} & 1 \cr}
\right)
\left( \matrix{
1 & 0 \cr
0 & - 1 \cr}
\right) 
\left( \matrix{
1 & \tau \cr
\tilde{\tau} & 1 \cr}
\right)^{- 1} \ . 
\label{14}
\ee
For this parametrization of $\sigma_R$, the integration measure
is~\cite{Zir96} the flat Berezinian $\prod_{p q} {\rm d} (\tau_{p q}
\tilde{\tau}_{q p})$, as is the case for $(Y, \tilde{Y})$. Complete
identity of the two saddle-point manifolds is then guaranteed if for
each set of block indices $(p, q)$ there exists a one-to-one map of
the two sets of matrices $(Y_{p q}, \tilde{Y}_{q p})$ and $(\tau_{p
  q}, \tilde{\tau}_{q p})$ onto each other. That follows from the
facts that all these supermatrices have dimension four, possess the
same symmetries including a compact parametrization of the
Fermion-Fermion block, and together parametrize the same supermanifold
(the extension of Efetov's coset space~\cite{Efe83} from the two-point
function to the $(P, Q)$ correlation function). It then follows that
all $(P, Q)$ correlation functions for time-reversal invariant graphs
and for the GOE pairwise coincide, both for closed and for open
systems.

{\it Discussion}. We have proved the BGS conjecture for quantum graphs
in its most general form both for closed and for open graphs in the
limit of infinite bond number $B$. The proof involves a number of
assumptions. (i) We have limited ourselves to graphs that are
time-reversal invariant (orthogonal symmetry). We expect, however,
that the proof can be straightforwardly extended to graphs that are
not time-reversal invariant (unitary symmetry). (ii) Graphs must have
incommensurate bond lengths. That assumption is essential as it allows
the average over the wave number $k$ to be replaced by averages over
the phases $\phi_b = k L_b$ and enables the use of the color-flavor
transformation. (iii) Graphs are completely connected. The removal of
a finite number of bonds probably does not affect our results for $B
\to \infty$. Otherwise, we expect qualitative changes that might be
caused, for instance, by Anderson localization. (iv) Graphs are
classically mixing. The ensuing condition on the spectrum of the PF
operator (existence of a gap separating the eigenvalue $+ 1$ from the
rest of the spectrum) guarantees that the contribution of the massive
modes to all $(P, Q)$ correlation functions vanishes for closed
graphs, and analogously for open graphs. In Refs.~\cite{Tan01, Gnu04,
  Gnu05, Gnu08, Gnu10} it is shown that weaker conditions on the
spectrum of the PF operator suffice to guarantee certain fluctuation
properties of the GOE type. It is not clear how such conditions relate
to conditions on the time evolution of the classical probability
density in directed bond space and, thus, to classical chaos.

In Refs.~\cite{Plu13a, Plu13b} the complete set of $(P, Q)$ $S$-matrix
correlation functions for graphs was calculated explicitly in the
Ericson regime, i.e., for $\sum_\alpha T^{(\alpha)} \gg 1$. It was
conjectured that these results are generic. The present paper confirms
that conjecture. Beyond that regime our results are only implicit. We
prove the identity of all $(P, Q)$ correlation functions for graphs
and for the GOE without being able to work out these functions
explicitly (except for $P = 1 = Q$).  

In Refs.~\cite{Aga95, And96a, And96b} a field-theoretical approach to
quantum chaos based upon the PF operator and on the non-linear sigma
model was advocated. Our work shows that the PF operator does indeed
determine essential features of the problem. Knowledge of that
operator is not sufficient, however. As shown below Eq.~(\ref{7}), the
masses of the modes $Z_{p b d s, q b d' s'}$ with $d \neq d'$ are
determined by quantum amplitudes that go beyond the classical PF
operator.

The authors are grateful for valuable comments to A. Altland,
P. Cejnar, S. Gnutzmann, J. Kvasil, and M. Zirnbauer. ZP acknowledges
support by the Czech Science Foundation under Project No P203 - 13 -
07117S.


\begin{thebibliography}{99}

\bibitem{Boh84}O. Bohigas, M. J. Giannoni, and C. Schmit, Phys. Rev.
Lett. {\bf 52}, 1 (1984).

\bibitem{McD79}S. W. McDonald and A. N. Kaufman, Phys. Rev. Lett.
{\bf 42}, 1189 (1979).

\bibitem{Cas80}G. Casati, F. Valz-Gris, and I. Guarneri, Lett. Nuovo
Cimento Soc. Ital. Fis. {\bf 28}, 279 (1980).

\bibitem{Ber81}M. V. Berry, Ann. Phys. (N.Y.) {\bf 131}, 161 (1981).

\bibitem{Dys62}F. J. Dyson, J. Math. Phys. {\bf 3}, 1199 (1962).

\bibitem{Haa10}F. Haake, {\it Quantum Signatures of Chaos}, 3rd
Edition, Springer-Verlag, Heidelberg/New York (2010).

\bibitem{Mue04}S. M{\"u}ller, S. Heusler, P. Braun, F. Haake, and A.
Altland, Phys. Rev. Lett. {\bf 93}, 014103 (2004).

\bibitem{Mue05}S. M{\"u}ller, S. Heusler, P. Braun, F. Haake, and A.
Altland, Phys. Rev. E {\bf 72}, 046207 (2005).

\bibitem{Heu07}S. Heusler, S. M{\"u}ller, A. Altland, P. Braun, and
F. Haake, Phys. Rev. Lett. {\bf 98}, 044103 (2007).

\bibitem{Heu09}S. Heusler, S. M{\"u}ller, A. Altland, P. Braun, and
F. Haake, New J. Phys. {\bf 11}, 103205 (2009).

\bibitem{Kot99}T. Kottos and U. Smilansky, Ann. Phys. (N.Y.) {\bf 274},
76 (1999).

\bibitem{Gnu04}S. Gnutzmann and A. Altland, Phys. Rev. Lett. {\bf 93},
194101 (2004).

\bibitem{Gnu05}S. Gnutzmann and A. Altland, Phys. Rev. E {\bf 72},
056215 (2005).

\bibitem{Plu13a}Z. Pluha{\v r} and H. A. Weidenm{\"u}ller,
  Phys. Rev. Lett. {\bf 110}, 034101 (2013). 

\bibitem{Plu13b}Z. Pluha{\v r} and H. A. Weidenm{\"u}ller, Phys.
Rev. E {\bf 88}, 022902 (2013).

\bibitem{Gnu06}S. Gnutzmann and U. Smilansky, Adv. Phys. {\bf 55},
527 (2006).

\bibitem{Bar01}F. Barra and P. Gaspard, Phys. Rev. E {\bf 63},
066215 (2001).

\bibitem{Pak01}P. Pakonski, K. Zyczkowski, and M. Kus, J. Phys. A
{\bf 34}, 9303 (2001).

\bibitem{Zir96}M. R. Zirnbauer, J. Phys. A: Math. Gen. {\bf 29},
7113 (1996).

\bibitem{Meh04}M. L. Mehta, {\it Random Matrices}, 3rd edition,
Academic Press, , New York (2004).

\bibitem{Efe83}K. B. Efetov, Adv. Phys. {\bf 32}, 53 (1983).

\bibitem{Ver85}J. J. M. Verbaarschot, H. A. Weidenm{\"u}ller, and
M. R. Zirnbauer, Phys. Rep. {\bf 129}, 367 (1985).

\bibitem{Ver85a}J. J. M. Verbaarschot and M. R. Zirnbauer, J.
Phys. A {\bf 18}, 1093 (1985).

\bibitem{Tan01}G. Tanner, J. Phys. A {\bf 34}, 8485 (2001).

\bibitem{Gnu08}S. Gnutzmann, J. P. Keating, and F. Piotet, Phys. Rev.
Lett. {\bf 101}, 264102 (2008).

\bibitem{Gnu10}S. Gnutzmann, J. P. Keating, and F. Piotet, Ann. Phys.
(N.Y.) {\bf 325}, 2595 (2010).

\bibitem{Aga95}O. Agam, B. L. Altshuler, and A. V. Andreev, Phys. Rev.
Lett. {\bf 75}, 4398 (1995).

\bibitem{And96a}A. V. Andreev, O. Agam, B. D. Simons, and B. L.
Altshuler, Phys. Rev. Lett. {\bf 76}, 3947 (1996).

\bibitem{And96b}A. V. Andreev, B. D. Simons, O. Agam, and B. L.
Altshuler, Nucl. Phys. B {\bf 482}, 536 (1996).

\end{thebibliography}
\end{document}